\documentclass[12pt]{article}
\usepackage{amssymb}
\usepackage{amsmath}
\usepackage{graphicx}
\newcommand{\be}{\begin{equation}}
\newcommand{\ee}{\end{equation}}
\newcommand{\bea}{\begin{eqnarray}}
\newcommand{\eea}{\end{eqnarray}}
\newcommand{\Dd}{{\cal{D}}}

\def\APJ{ Astroph.~J.~}

\newcommand{\ead}[1]{\vspace*{5pt} {E-mail: \tt{#1}}}

\begin{document}

\title{
\vskip -50pt
{\begin{normalsize}
\mbox{} \hfill DAMTP-2007-108\\
\vskip 50pt
\end{normalsize}}
Dynamics in nonlocal linear models in the
Friedmann--Robertson--Walker metric }

\author{I.~Ya.~Aref'eva\\
Steklov Mathematical Institute, Russian Academy of Sciences,\\
Gubkina st. 8, 119991, Moscow, Russia\\
\ead{arefeva@mi.ras.ru}\\[7.2mm]
L.~V.~Joukovskaya\\
DAMTP, Centre for Mathematical Sciences, University of
Cambridge,\\
Wilberforce Road, CB3 0WA, Cambridge, UK\\
\ead{l.joukovskaya@damtp.cam.ac.uk}\\[7.2mm]
S.~Yu.~Vernov\\
Skobeltsyn Institute of Nuclear Physics, Moscow State University,\\
Vorobyevy Gory, 119991, Moscow, Russia\\
\ead{svernov@theory.sinp.msu.ru}}
\date{}
\maketitle

\begin{abstract}
A general class of cosmological models driven by a nonlocal scalar
field inspired by the string field theory is studied. Using the
fact that the considering linear nonlocal model is equivalent to
an infinite number of local models we have found an exact special
solution of the nonlocal Friedmann equations. This solution
describes a monotonically increasing Universe with the phantom
dark energy.
\end{abstract}



\section{Introduction}

Recently string theory and brane cosmology have been intensively
discussed as promising candidates for the theoretical explanation
of the obtained experimental data  (see for example
\cite{string-cosmo}--\cite{Biswas}).

The purpose of this paper is to present new results concerning
studies of nonlocal linear models in the
Friedmann--Robertson--Walker Universe. These models are inspired
by the string field theory (SFT) (for review of the SFT
see~\cite{review-sft}). A distinguished feature of nonlocal linear
and nonlinear models~\cite{IA1}--\cite{Jukovskaya0710} is the
presence of infinite number of higher derivative terms (note also
nonlocal models in the Minkowski
space-time~\cite{Woodard}--\cite{LJ}). For special values of the
parameters these models describe linear approximations to the
cubic bosonic or nonBPS fermionic SFT nonlocal tachyon models,
p-adic string models or the models with the invariance of the
action under the shift of the dilaton field to a constant. The
NonBPS fermionic string field tachyon nonlocal model has been
considered as a candidate for the dark energy~\cite{IA1}.

Present cosmological observations \cite{cosmo-obser} do not
exclude an evolving dark energy (DE) state parameter $w$, whose
current value can be less than  $-1$, that means the violation of
the null energy condition (NEC) (see \cite{DE1,DE2} for a review
of the DE problems and \cite{0312430} for a search  for a
super-acceleration phase of the Universe).

Field theories, which violate the
NEC~\cite{Hawking-Ellis,Venziano}, are of interest not only for
the construction of cosmological dark energy models with the state
parameter $w<-1$, but also for the solution of the cosmological
singularity problem. A possible way to avoid cosmological
singularities consists of dealing with nonsingular bouncing
cosmological solutions. In this scenario the Universe contracts
before the bounce~\cite{Turok}. Such models have strong coupling
and higher-order string corrections are inevitable. It is
important to construct nonsingular bouncing cosmological solutions
in order to make a concrete prediction of bouncing cosmology.

A simple possibility to violate the NEC is just to deal with a
phantom field. In the present paper we consider nonlocal models
which are linear and admit solutions, which are linear
combinations of local fields. Some of these local fields are
phantoms. Namely due to the presence of these ghost excitations
such nonlocal models present an interest for cosmology.

At the same time there are well known problems with instability of
quantum models with phantoms, namely a lost of unitarity and so
on. We believe that nonlocal SFT models in true vacua are stable
with respect to quantum fluctuations. This question has to be
consider in the full SFT framework and demands further
investigations. We also believe that due to these string theory
origin the corresponding nonlocal cosmological models, which are
nonlinear in matter fields, have no problem with instability in
the quantum case. In this paper we consider only the classical
case and models, which are linear in a nonlocal scalar field.

 In our previous paper~\cite{AJV0701184} as well as in
paper~\cite{AVzeta} nonlocal linear models already have been
studied. In~\cite{AJV0701184} nonlocal linear model has been
studied in the flat space-time and we have proposed special
deformations of the potential, which allows us to get the same
scalar field solutions in flat and nonflat (the FRW metric) cases.
As result we have obtained nonlinear models in the FRW metric.
In~\cite{AVzeta} few exact solutions to linear model in the FRW
metric have been found. In this paper we present a systematic
method that permits us to transform the initial nonlocal system
into infinity set of local systems. The choice of a local system
is equivalent to the choice of a special solution of the nonlocal
system. This approach allows us to use the standard method of
analysis of the differential equations and in particular to find
exact solutions.

The paper is organized as follows. In Section 2 we describe string
inspired models with quadratic nonlocal potentials. In Section 3
we assume that the metric is given and consider the equation of
motion as an equation for the nonlocal scalar field. We construct
solutions, using eigenfunctions of the $\square_g$-operator with
eigenvalues, belonging to the set of roots of the characteristic
equation. In Section 4 we find values of the energy density and
pressure for these solutions. In Section 5 we consider the
Friedmann--Robertson--Walker Universe and find local models, which
correspond to particular solutions of the initial nonlocal model.
In the case of dilaton massless scalar field we construct the
general solutions for the corresponding local model, which are the
special exact solution for the initial nonlocal model as well. We
analyze cosmological properties of the obtained solutions.

\section{Nonlocal linear models}

In this paper we consider a model of gravity coupling with a
nonlocal scalar field,  which induced by string field theory
 \begin{equation}
\label{ACTION} S=\int d^4x\sqrt{-g}\left(\frac{M_p^2}{2}R+
\frac{M_s^4}{g_4}\left(\frac{1}{2}\phi\,F\left(-\square_g/M_s^2\right)\phi
-\Lambda^\prime \right)\right),
\end{equation}
where $g_{\mu\nu}$ is the metric tensor (we use the signature
$(-,+,+,+)$ ),
$\square_g=\frac1{\sqrt{-g}}\partial_{\mu}\sqrt{-g}g^{\mu\nu}\partial_{\nu}$,
$M_p$
 is  a mass Planck, $M_s$ is a characteristic string scale
 related with the  string tension $ \alpha^{\prime}$:
$M_s=1/\sqrt{ \alpha^{\prime}}$, $\phi$ is a dimensionless scalar
field, $g_4$ is a dimensionless four dimensional effective
coupling constant related with the ten dimensional  string
coupling constant $g_0$ and the compactification scale.
$\Lambda=\frac{M_s^4}{g_4}\Lambda^\prime $ is an effective four
dimensional cosmological constant.

The form of the function $F$ is inspired by a nonlocal action
appeared in the string field theory. We consider the case
\begin{equation}
\label{F} F(z)=-\xi^2 z+1-c\:e^{-2z},
\end{equation}
where $\xi$ is a real parameter and $c$ is a positive constant.
Using dimensionless space-time variables and a rescaling we can
rewrite (\ref{ACTION}) for $F$ given by (\ref{F}) as follows
\begin{equation}
\label{action-1}  S=\int
d^4x\sqrt{-g}\left(\frac{m_p^2}{2}R+\frac{\xi^2 }{2}
\phi\,\square_g\phi+
 \frac{1}{2}\left(\phi^2-c\:\Phi^2\right)-\Lambda^\prime\right),
\end{equation}
where
\begin{equation*}
\Phi=e^{\square_g}\phi
\end{equation*}
and $m_p^2=g_4M_p^2/M_s^2$. Generally speaking the string scale
does not coincide with the Plank mass. That gives a possibility to
get a realistic value of $\Lambda$.

The form of the term $(e^{\square_g}\phi)^2$
 is analogous to the form of the
interaction term for the tachyon field in the SFT action. The case
of the open cubic superstring field theory tachyon corresponds to
$\xi^2={}-1/\left(4\ln\left(\frac{4}{3\sqrt{3}}\right)\right)\approx
0.9556$ and $c=3$ (see \cite{AJK}-\cite{Trento}).

The equation of motion for the scalar field has the following form
\begin{equation}
\label{equphiNF} (\xi^2\square_g +1)e^{-2 \square_g}\phi= c\:
\phi.
\end{equation}

The energy-momentum tensor
\begin{equation}\label{Tmunu}
T_{\alpha\beta}={}-\frac{2}{\sqrt{-g}}\frac{\delta{S}}{\delta
g^{\alpha\beta}}
\end{equation}
has the following explicit form
\begin{equation*}
T_{\alpha\beta}=-g_{\alpha\beta}\left(\frac{1}{2}\phi^2-
\frac{\xi^2}{2}\partial_{\mu}\phi
\partial^{\mu}{\phi}-\frac{c}{2}(e^{\square_g}\phi)^2-\Lambda^{\prime}\right)-
\xi^2 \partial_{\alpha}\phi \partial_{\beta}{\phi}-{}
\end{equation*}
\begin{equation*}
{}-g_{\alpha\beta}\, c \int\limits_0^1
d\rho\left[(e^{(1+\rho)\square_g}\phi)(\square_g
e^{(1-\rho)\square_g}\phi)
+(\partial_{\mu}e^{(1+\rho)\square_g}\phi)(\partial^{\mu}e^{
(1-\rho)\square_g}\phi)\right] +{}
\end{equation*}
\begin{equation*}
{}+2 c\int\limits_{0}^{1} d\rho
\left(\partial_{\alpha}e^{(1+\rho)\square_g}\phi\right)\left(\partial_{\beta}
e^{(1-\rho)\square_g}\phi\right).
\end{equation*}
Note that the energy-momentum tensor $T_{\alpha\beta}$ includes
the nonlocal terms, so the Einstein's equations are nonlocal ones.

\section{Generalization of Flat Dynamics}

\subsection{Flat Dynamics}
In the flat case action (\ref{ACTION}) has the following form:
\begin{equation}
\label{action-1flat} S_{flat}=\frac{1}{2}\int d^4x \phi
F(-\square)\phi.
\end{equation}
If the scalar field $\phi$ depends only on time, then equation of
motion (\ref{equphiNF}) is reduced to the following linear
equation:
\begin{equation}
\label{F-eq} F(\partial_0^2)\phi(t)=0.
\end{equation}

A plane wave $\phi=e^{\alpha t}$ is a solution of (\ref{F-eq}) if
$\alpha$ is a root of the characteristic equation
\begin{equation}
F(\alpha^2)=0. \label{ch-F}
\end{equation}

For a case of $F$ given by (\ref{F}) equation (\ref{F-eq}) has the
following form
\begin{equation}
-\xi^2\partial_0^2 \phi + \phi- c\:e^{-2\partial_0^2}\phi=0
\label{1c}.
\end{equation}
This equation has been analysed in detail in our
paper~\cite{AJV0701184}. Using the explicit form the function
$\phi(t)$ we have found the solutions of equations of motion and
the corresponding values of the energy density and pressure. In
this paper we generalise these calculations for non-flat case.

\subsection{The equation of motion in an arbitrary metric}

Let us consider eq.~(\ref{equphiNF}). Really this equation is a
consequence of the Einstein's equations, hence, both the metric
$g_{\mu\nu}$ and the scalar field $\phi$ are unknown. We assume
that the metric $g_{\mu\nu}$ is given and consider
eq.~(\ref{equphiNF}) as an equation in $\phi$.

In this paper we study solutions in the following form:
\begin{equation}
\label{phiexpsum}
    \phi=\sum_{n=1}^{N} \phi_n,
\end{equation}
where $N$ is a natural number, $\phi_n$ is a solution of the
following equation:
\begin{equation}
\label{even} \square_g \phi_n ={}- \alpha_n^2 \phi_n,
\end{equation}
and $\alpha_n$ are solutions to the corresponding characteristic
equation:
\begin{equation}
F(\alpha_n^2)\equiv-\xi^2\alpha_n^2  + 1 - c\:e^{-2\alpha_n^2}=0.
\label{5c}
\end{equation}

Without loss of generality we can assume that for any $n$ and
$k\neq n$ the conditions $\alpha^2_n\neq\alpha^2_k$ are satisfied.
Indeed, if the sum (\ref{phiexpsum}) includes two summands
$\phi_{k_{1}}$ and $\phi_{k_{2}}$, which correspond to one and the
same $\alpha_k^2$, then we can consider them as one summand
$\phi_k\equiv \phi_{k_{1}}+\phi_{k_{2}}$, which corresponds to
$\alpha_k^2$.

We start with construction of a solution to equation
(\ref{equphiNF}) in the case $N=1$:
\begin{equation}
\label{eve} \square_g \phi ={}- \alpha^2 \phi,
\end{equation}
where $\alpha$ is a root of~(\ref{5c}). Note that this ansatz is
widely used in studying of nonlocal linear models
\cite{Biswas,Cline,AVzeta,Lidsey07,Calcagni07,Mulryne}. Equation
(\ref{5c}) has the following solutions
\begin{equation}
\label{sol-lamb-k} \alpha_n={}\pm\frac{1}{2\xi}\sqrt{4+ 2\xi^2
W_n\left({}-\frac{2c\:e^{-2/\xi^2}}{\xi^2}\right)}, \quad n=0,\pm
1,\pm 2,...
\end{equation}
where $W_n$ is  the n-s branch of the  Lambert function satisfying
a relation $W(z)e^{W(z)}=z$. The Lambert function is a multivalued
function, so eq.~(\ref{5c}) has an infinite number of roots.
Parameters $\xi$ and $c$ are real, therefore if $\alpha_n$ is a
root of (\ref{5c}), then the adjoined number $\alpha_n^*$ is a
root as well. Note that  if $\alpha_n$ is a root of (\ref{5c}),
then $-\alpha_n$ is a root too.

 If $\alpha^2=\alpha^2_0$ is a multiple root, then at this point
$F(\alpha^2_0)=0$ and $F'(\alpha^2_0)=0$. These equations give
that
\begin{equation}
\label{z-0} \alpha _0^2=\frac{1}{\xi^2}-\frac12,
\end{equation}
 hence $\alpha^2_0$ is a real number and all multiple roots of
 $F(\alpha^2_0)=0$ are either real or pure imaginary.
Double roots exist if and only if
\begin{equation}
\label{c-xi} c=\frac{\xi^2}{2}\:e^{(2/\xi^2-1)}.
\end{equation}

Note that the existence of double roots means that there exist
solutions of equation~(\ref{equphiNF}), which does not satisfy of
equation (\ref{eve}), but satisfy the following equation
\begin{equation}
\label{eve2} (\square_g+\alpha^2)(\square_g+\alpha^2) \phi =0.
\end{equation}

In the flat case an example of such solution is the function
$\phi(t)=t\exp(\alpha t)$ (see~\cite{AJV0701184}).
 All roots for any
$\xi$ and $c$ are no more than double degenerated, because
$F''(\alpha^2_0)\neq 0$. In this paper we consider such values of
$\xi$ and $c$ that equality (\ref{c-xi}) is not satisfied and all
roots are simple ones. Under this assumption we can consider the
set of the solutions (\ref{phiexpsum}) as a quite general
solution.

\subsection{Real Roots of the Characteristic Equation}

For some values of the parameters $\xi$ and $c$ eq.~(\ref{5c}) has
real roots. To mark out real values of $\alpha$ we will denote
real $\alpha$ as $m$: $m=\alpha$.

\begin{figure}[h]
\centering
\includegraphics[width=47.2mm]{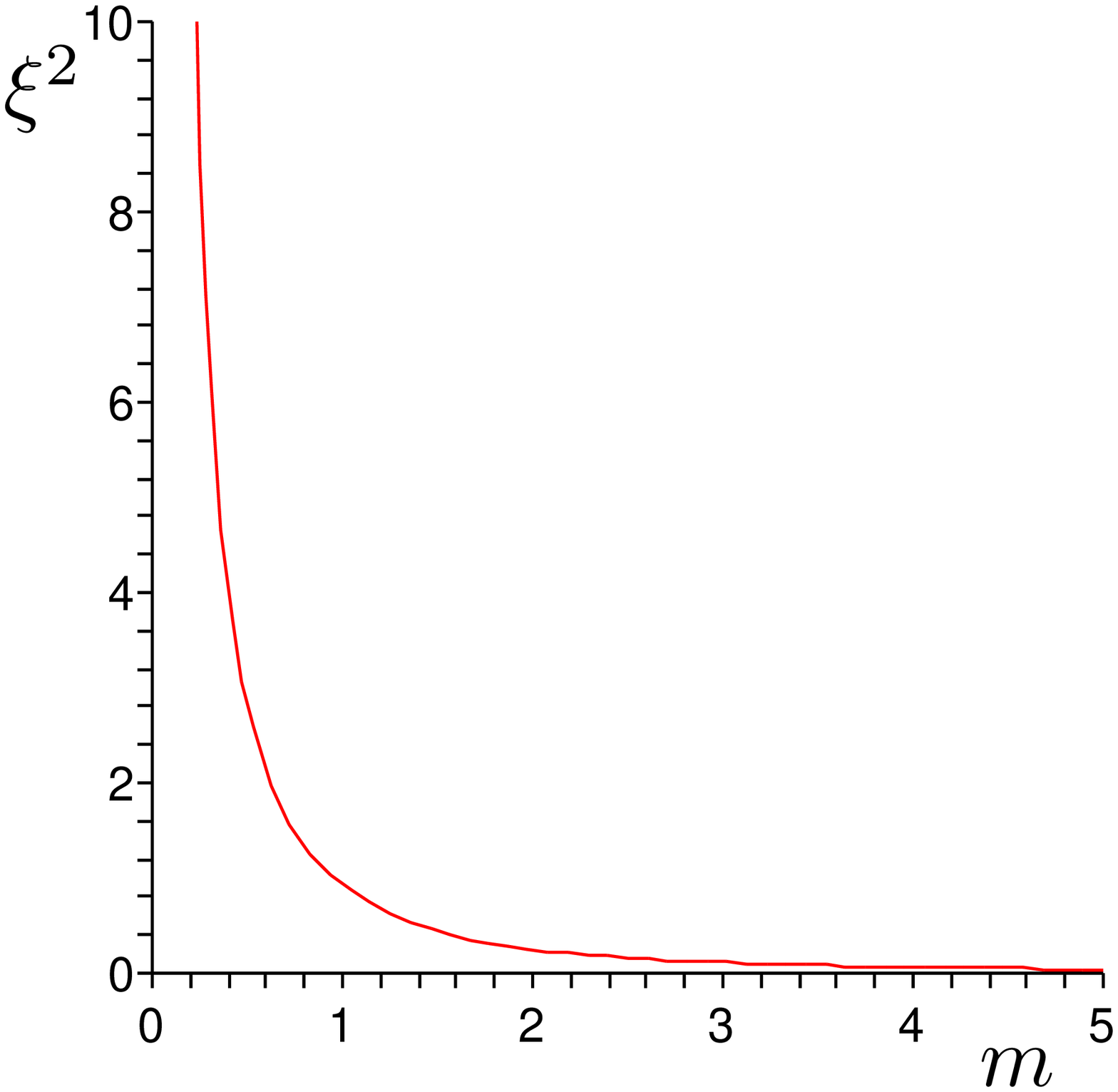} \ \ \ \ \
\includegraphics[width=47.2mm]{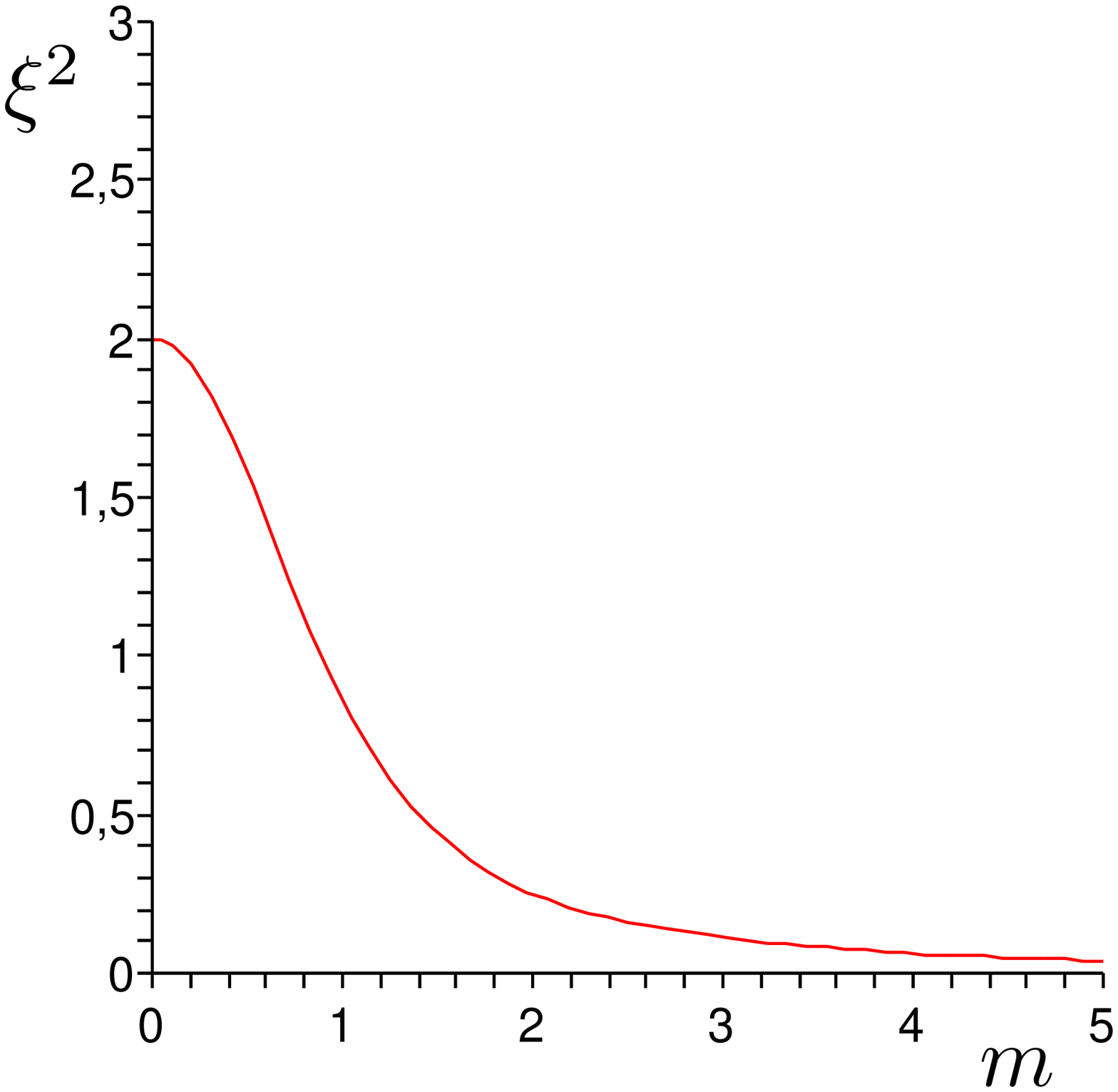} \ \ \ \ \
\includegraphics[width=47.2mm]{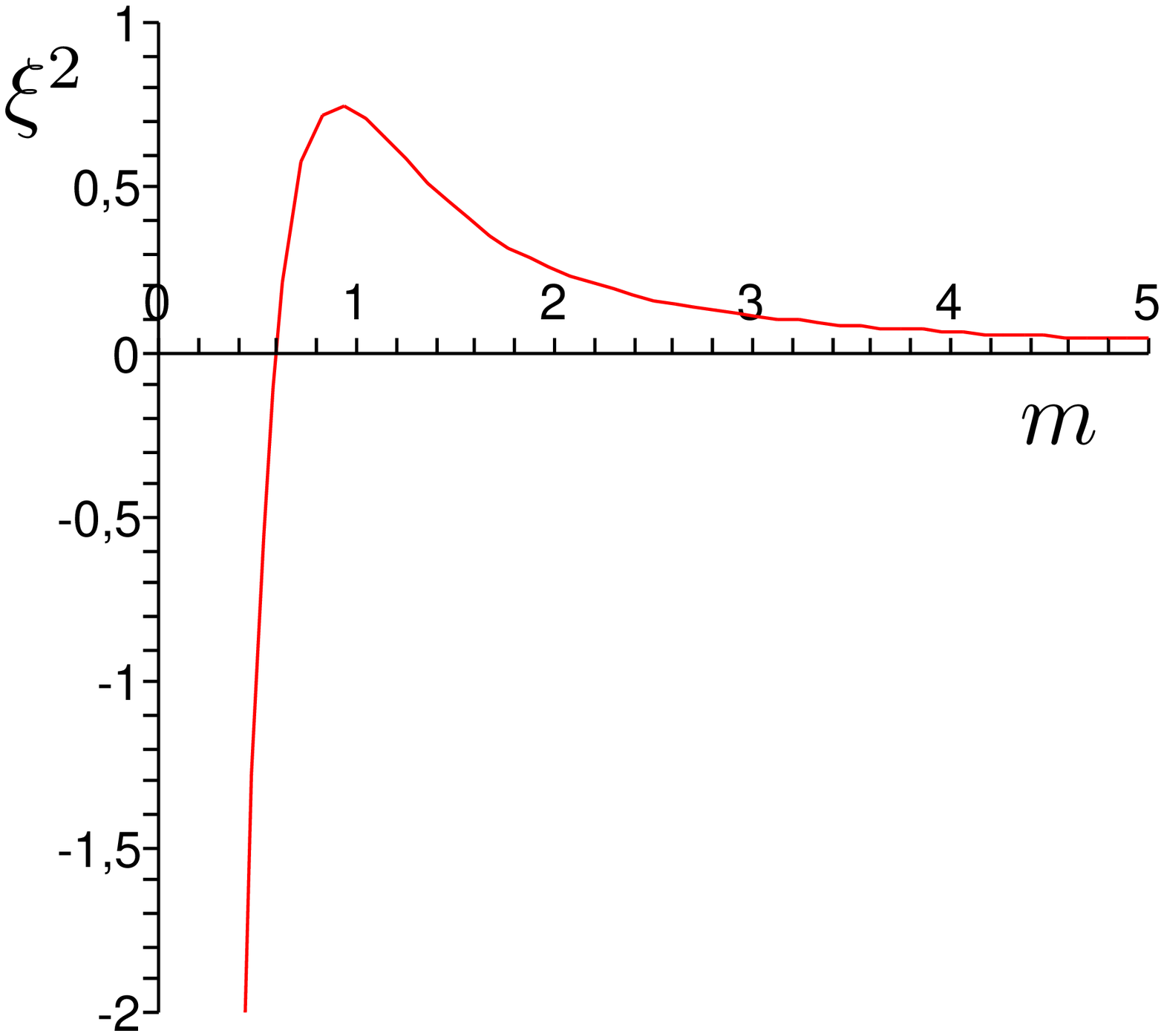}
\caption{ The dependence of the function $g(m^2,c)$, which is
equal to $\xi^2$, on $m$ at $c=1/2$ (left), $c=1$ (center) and
$c=2$ (right). } \label{xi2_m}
\end{figure}

To determine values of the parameters at which eq.~(\ref{5c}) has
real roots we rewrite this equation in the following form:
\begin{equation}\label{equxi}
    \xi^2 = g(m^2,c), \quad {\mbox {where}} \quad
    g(m^2,c)=\frac{e^{2m^2}-c}{m^2e^{2m^2}}.
\end{equation}
The dependence of $g(m,c)$ on $m$ for different $c$ is presented
in Fig.~(\ref{xi2_m}). This function has a maximum at $m_{max}^2$
\begin{equation}
m_{max}^2=-\frac12-\frac12 W_{-1}\left(-\frac{e^{-1}}{c}\right),
\label{max}
\end{equation}
provided $c$ is such that
$W_{-1}\left(-\frac{e^{-1}}{c}\right)<-1$, in the other words
$0<c<1$.

There are three different cases (see Fig.~\ref{xi2_m}).
\begin{itemize}
\item If $c<1$, then eq.~(\ref{5c}) has two simple real roots:
$m=\pm m_1$ for any values $\xi$.

\item If $c=1$, then eq.~(\ref{5c}) has a zero root. Nonzero real
roots exist if and only if $\xi^2<2$.

\item If $c>1$, then eq.~(\ref{5c}) has
\begin{itemize}
\item no real roots for $\xi^2>\xi^2_{max}$, where
\begin{equation}
\label{xi-max} \xi_{max}^2=\frac{1-ce^{-2m_{max}^2}}{m_{max}^2}
={}-\frac{2}{W_{-1}(-e^{-1}/c)}
\end{equation}

\item two real double roots $m=\pm m_{max}$ for
$\xi^2=\xi^2_{max}$

\item four real simple roots for  $\xi^2<\xi^2_{max}$. In this
case we have the following restriction on real roots:
$m^2>\frac12\ln c$.
\end{itemize}
\end{itemize}

Note that values of roots do not depend on $H(t)$ and, therefore,
coincide with roots in the flat case, which have been found
in~\cite{AJV0701184}.

\section{Energy Density and Pressure}
\subsection{General Formula} Let us calculate the energy density
and the pressure for the solution (\ref{phiexpsum}). Up to this
moment we do not put any restrictions on the metric tensor
$g_{\mu\nu}$, now we start to consider the case of the spatially
flat Friedmann--Robertson--Walker  Universe:
\begin{equation}
\label{mFr} ds^2={}-dt^2+a^2(t)\left(dx_1^2+dx_2^2+dx_3^2\right)
\end{equation}
and spatially homogeneous solutions $\phi(t)$.  In this case
\begin{equation}
T_{\alpha\beta}=g_{\alpha\beta}\,\mbox{diag}\{{\cal E},{-\cal
P},{-\cal P},{-\cal P}\},
\end{equation}
where the energy density ${\cal E}$ and pressure ${\cal P}$ are as
follows
\begin{equation}
\label{E-phi-H-p}
 {\cal E}={\cal E}_k+{\cal E}_p+{\cal E}_{nl2}+{\cal
 E}_{nl1}+\Lambda^\prime,\qquad
{\cal P}={\cal E}_k-{\cal E}_p+{\cal E}_{nl2}-{\cal
E}_{nl1}-\Lambda^\prime.
\end{equation}
Nonlocal term ${\cal E}_{nl1}$ plays a role of an extra potential
term and ${\cal E}_{nl12}$ plays a role of an extra kinetic term.
The explicit form of the terms in the R.H.S. of (\ref{E-phi-H-p})
is~\cite{Yang,LJ} as follows

\begin{equation}
\label{El-Enl}
\begin{array}{ll}
\displaystyle {\cal E}_k&=\displaystyle \frac{\xi^2}{2}(\partial_0\phi)^2,\\[2.7mm]
 \displaystyle {\cal E}_p&=\displaystyle{}
 -\frac{1}{2}\left(\phi^2-c(e^{\cal D}\phi)^{2}\right),\\[2.7mm]
\displaystyle{\cal E}_{nl1}&=\displaystyle
c\int\limits_{0}^{1}\left(e^{(1+\rho) {\cal D}} \phi\right)
 \left({}-{\cal D} e^{(1-\rho){\cal D}}  \phi\right) d \rho,
\\[2.7mm]
\displaystyle{\cal
 E}_{nl2}&=\displaystyle{} -c\int\limits_{0}^{1}\left(\partial
e^{(1+\rho){\cal D}} \phi\right) \left(\partial e^{(1-\rho) {\cal
D}}\phi\right) d \rho,
\end{array}
\end{equation}
where
\begin{equation}
\label{cal-D} {\cal D}\equiv {}- \partial_0^2- 3H(t)\partial_0,
\quad  H=\frac{\partial_0{a}}{a}.
\end{equation}

For $N=1$ we obtain
\begin{equation}\label{E1}
    {\cal E}\equiv E(\phi_1)+\Lambda^\prime=\frac{\eta_{\alpha_{1^{\vphantom {27}}}}}{2}\left(\left(\partial_0\phi_1\right)^2-
    \alpha_1^2\phi_1^2\right)+\Lambda^\prime,
\end{equation}
\begin{equation}\label{P1}
    {\cal P}\equiv P(\phi_1)-\Lambda^\prime =\frac{\eta_{\alpha_{1^{\vphantom {27}}}}}{2}\left(
    \left(\partial_0\phi_1\right)^2+
    \alpha_1^2\phi_1^2\right)-\Lambda^\prime,
\end{equation}
where for arbitrary $\alpha$
\begin{equation}
     \eta^{\vphantom{27}}_{\alpha}\equiv \xi^2+2\xi^2{\alpha}^2-2.
     \label{eta}
\end{equation}

Note that considering the flat space-time~\cite{AJV0701184}, we
have introduced the parameter $p_\alpha\equiv
\alpha^2\eta^{\vphantom{27}}_{\alpha}$. The use of parameter
$\eta^{\vphantom{27}}_{\alpha}$ instead of $p_\alpha$ is more
convenient, because we do not need to consider the case $\alpha=0$
separately.

 Hereafter we denote the energy density and pressure of
function $\phi(t)$ as the functionals $E(\phi)$ and $P(\phi)$,
respectively.

For the solution $\phi(t)=\phi_1(t)+\phi_2(t)$ it is convenient to
write the energy density in the following form
\begin{equation*}
    {\cal E}=E(\phi_1+\phi_2)+\Lambda^\prime=E(\phi_1)+E(\phi_2)
    +E_{cross}(\phi_1,\phi_2)+\Lambda^\prime,
\end{equation*}
where the functional $E_{cross}(\phi_1,\phi_2)$ is defined as
follows:
\begin{equation}
E_{cross}(\phi_1,\phi_2)=E_{k_{cr}}+E_{nl2_{cr}}+E_{p_{cr}}+E_{nl1_{cr}},
\end{equation}
\begin{equation*}
E_{k_{cr}}\equiv \xi^2\partial_0\phi_1\partial_0\phi_2,\qquad
E_{p_{cr}}\equiv{}
-\phi_1\phi_2+c\:e^{-\alpha_1^2-\alpha_2^2}\phi_1\phi_2,
\end{equation*}
\begin{equation*}
E_{nl1_{cr}}\equiv{}-c\!\int\limits_0^1\!\left[\left(
e^{(1+\rho){\cal D}}\phi_1\right){\cal D}\left( e^{(1-\rho){\cal
D}}\phi_2\right)+\left( e^{(1+\rho){\cal D}}\phi_2\right){\cal
D}\left( e^{(1-\rho){\cal D}}\phi_1\right)\right] d\rho,
\end{equation*}
\begin{equation*}
E_{nl2_{cr}}\equiv{} -c\!\int\limits_0^1\!\Bigl[\partial_0\left(
e^{(1+\rho){\cal D}}\phi_1\right)\partial_0\left( e^{(1-\rho){\cal
D}}\phi_2\right)+\partial_0\left( e^{(1+\rho){\cal
D}}\phi_2\right)\partial_0\left( e^{(1-\rho){\cal
D}}\phi_1\right)\Bigr] d\rho.
\end{equation*}

Using (\ref{5c}), we calculate $E_{nl2_{cr}}$:
\begin{equation}
    E_{nl2_{cr}}={}-\frac{c\left(e^{-2\alpha_1^2}
    -e^{-2\alpha_2^2}\right)}{\alpha_2^2-\alpha_1^2}\partial_0\phi_1\partial_0\phi_2={}
    -\xi^2\partial_0\phi_1\partial_0\phi_2.
\end{equation}
So
\begin{equation}
\label{ecrosskin} E_{nl2_{cr}}+E_{k_{cr}}=0.
\end{equation}
The straightforward calculation also gives that
\begin{equation}
E_{nl1_{cr}}={}-c\:e^{-\alpha_1^2-\alpha_2^2}\phi_1\phi_2+\frac{c
\left(\alpha_2^2e^{-2\alpha_1^2}
    -\alpha_1^2e^{-2\alpha_2^2}\right)}{\alpha_2^2-\alpha_1^2}\phi_1\phi_2=
    {}-E_{p_{cr}}.
\end{equation}

Therefore, we obtain that
\begin{equation}
E_{cross}(\phi_1,\phi_2)=0,\qquad\mbox{and}\qquad
P_{cross}(\phi_1,\phi_2)=0,
\end{equation}
where
\begin{equation}
P_{cross}(\phi_1,\phi_2)\equiv
E_{k_{cr}}+E_{nl2_{cr}}-E_{p_{cr}}-E_{nl1_{cr}}.
\end{equation}
So,
\begin{equation}
    E(\phi_1+\phi_2)=E(\phi_1)+E(\phi_2),
\end{equation}
\begin{equation}
    P(\phi_1+\phi_2)=P(\phi_1)+P(\phi_2).
\end{equation}
Finally, for the case of N summands we obtain (compare
with~\cite{Koshelev07,AJV0701184,Lidsey07}):
\begin{equation}
\label{E-gen}
    {\cal E}=E\left(\sum_{n=1}^{N} \phi_n\right)
    +\Lambda^\prime=\sum_{n=1}^{N} E(\phi_n)+\Lambda^\prime,
\end{equation}
\begin{equation}
\label{P-gen}
    {\cal P}=P\left(\sum_{n=1}^{N} \phi_n\right)
    -\Lambda^\prime=\sum_{n=1}^{N} P(\phi_n)-\Lambda^\prime.
\end{equation}

From formulas (\ref{E-gen}) and (\ref{P-gen}) we see that the
energy density and the pressure are sums of "individual" energy
densities and pressures respectively and have no crossing term.

In the case of an arbitrary metric $g_{\alpha\beta}$ and a scalar
field $\phi_n(t,x_1,x_2,x_3)$, which satisfies eq.~(\ref{even}),
we obtain that
\begin{equation*}
T_{\alpha\beta}(\phi_n)=-g_{\alpha\beta}\left(\frac{1}{2}\phi_n^2-
\frac{\xi^2}{2}\partial_{\mu}\phi_n
\partial^{\mu}{\phi_n}-\frac{c}{2}(e^{\square_g}\phi_n)^2\right)-
\xi^2 \partial_{\alpha}\phi_n\partial_{\beta}{\phi_n}-{}
\end{equation*}
\begin{equation*}
- c\:g_{\alpha\beta}\! \int\limits_0^1\!
d\rho\left[(e^{(1+\rho)\square_g}\phi_n)(\square_g
e^{(1-\rho)\square_g}\phi_n)
+(\partial_{\mu}e^{(1+\rho)\square_g}\phi_n)(\partial^{\mu}e^{
(1-\rho)\square_g}\phi_n)\right]+\!
\end{equation*}
\begin{equation*}
{}+2 c\int\limits_{0}^{1} d \rho
(\partial_{\alpha}e^{(1+\rho)\square_g}\phi_n)(\partial_{\beta}
e^{(1-\rho)\square_g}\phi_n)={}
\end{equation*}
\begin{equation*}
{}=g_{\alpha\beta}\left(\frac{\eta^{\vphantom{27}}_{\alpha_n}}{2}
\partial_{\mu}\phi_n\partial^{\mu}\phi_n
-\frac{\eta^{\vphantom{27}}_{\alpha_n}\alpha_n^2}{2}\phi_n^2\right)-\eta^{\vphantom{27}}_{\alpha_n}
\partial_{\alpha}\phi_n\partial_{\beta}\phi_n.
\end{equation*}

The energy-momentum tensor, which corresponds to the function
(\ref{phiexpsum}), is as follows
\begin{equation}
T_{\alpha\beta}=T_{\alpha\beta} \left(\sum_{n=1}^N
\phi_n\right)+g_{\alpha\beta}\Lambda^{\prime}=\sum_{n=1}^N
T_{\alpha\beta}(\phi_n)+g_{\alpha\beta}\Lambda^{\prime}.
\end{equation}

\subsection{Energy Density and Pressure for real $\alpha$}
As we have seen in Section 3 for some values of parameters $\xi$
and $c$ eq.~(\ref{5c}) has real roots. We denote as $\eta_m$ the
value of $\eta_\alpha$ for real $\alpha=m$:
\begin{equation}
 \eta_m=\xi^2\left(1+2m^2\right)-2
 =\frac{e^{2m^2}-c}{m^2e^{2m^2}}\left(1+2m^2\right)-2.
\end{equation}

\begin{figure}[h]
\centering
\includegraphics[width=47.2mm]{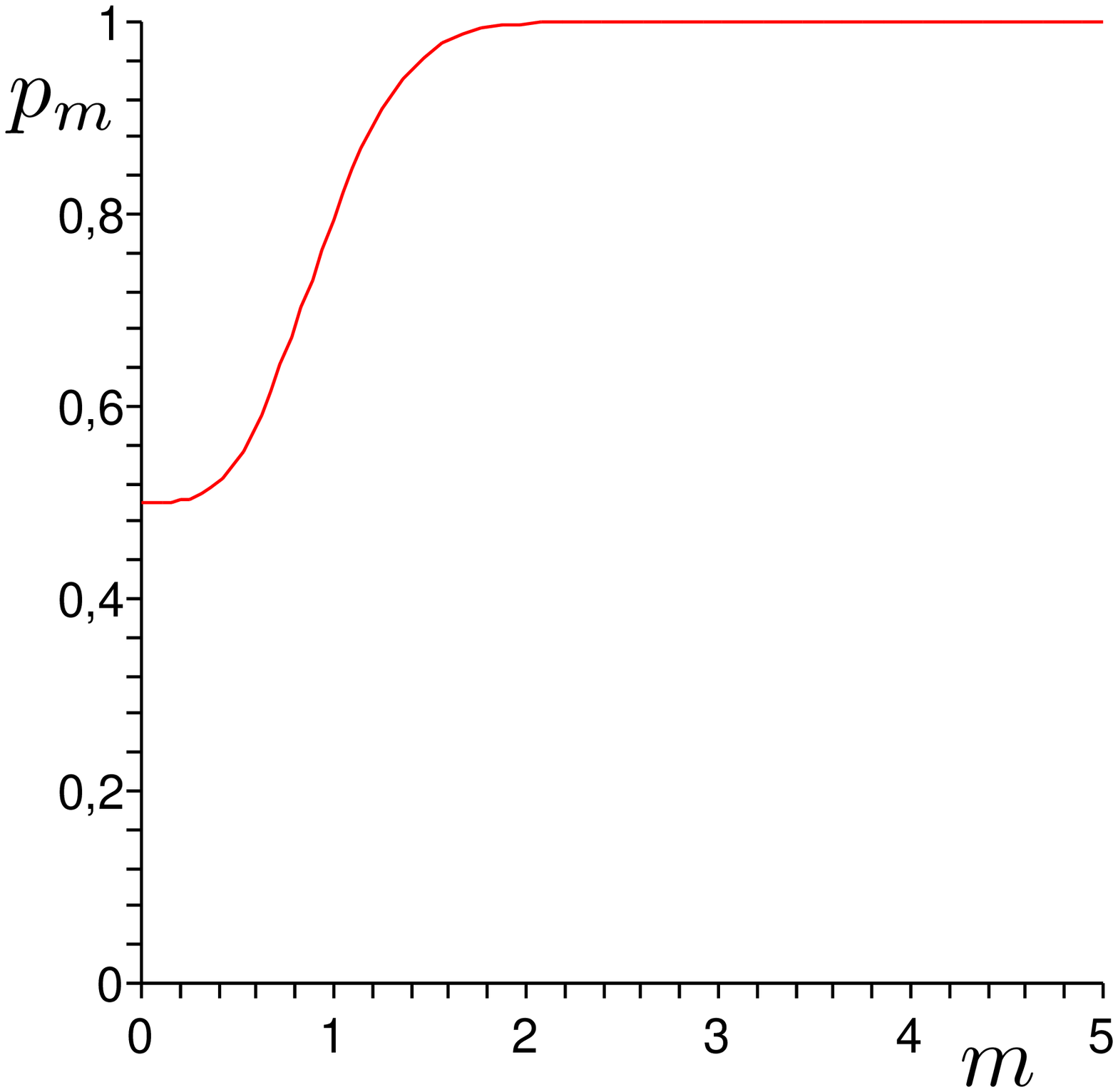} \ \ \ \ \
\includegraphics[width=47.2mm]{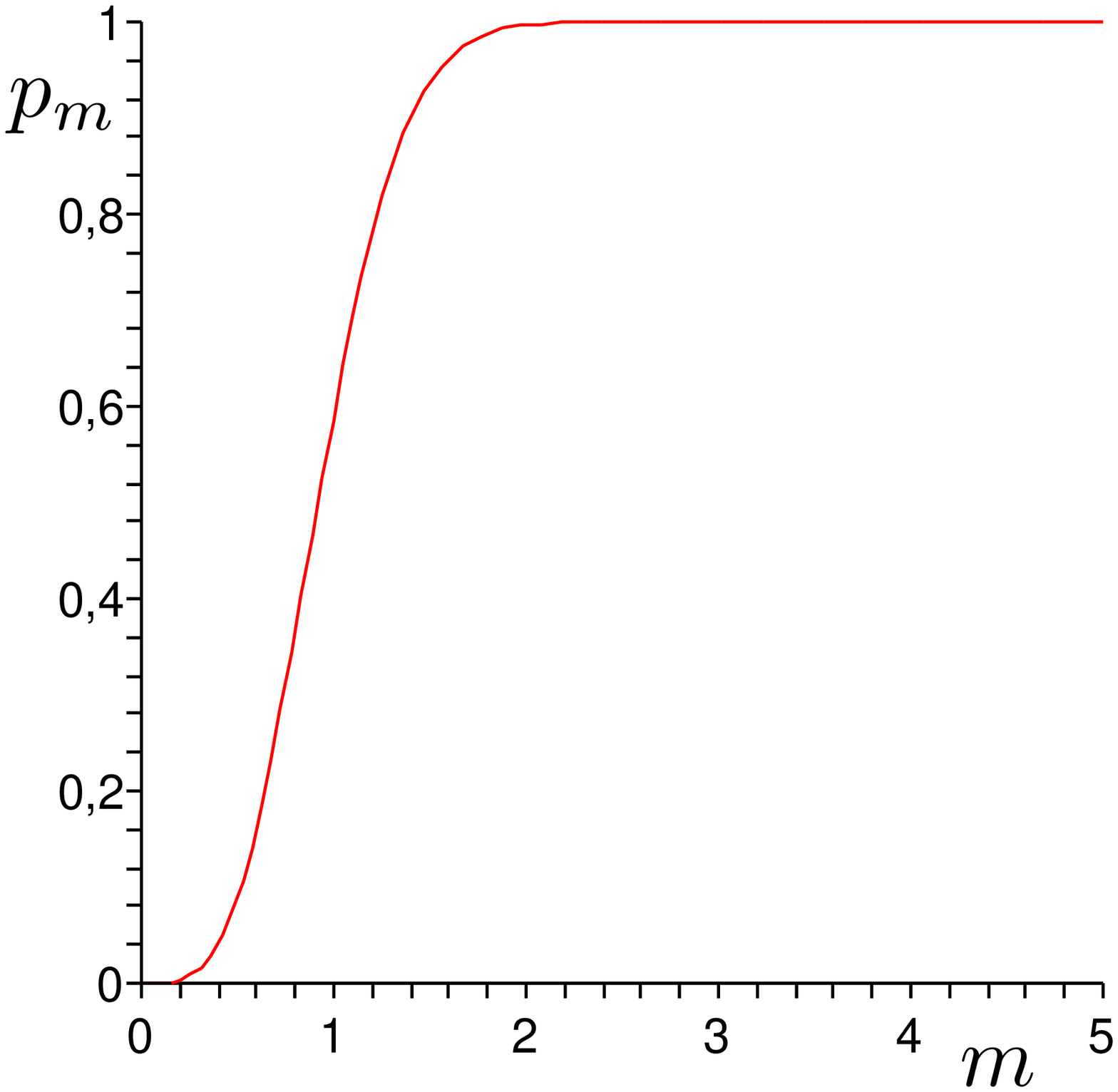} \ \ \ \ \
\includegraphics[width=47.2mm]{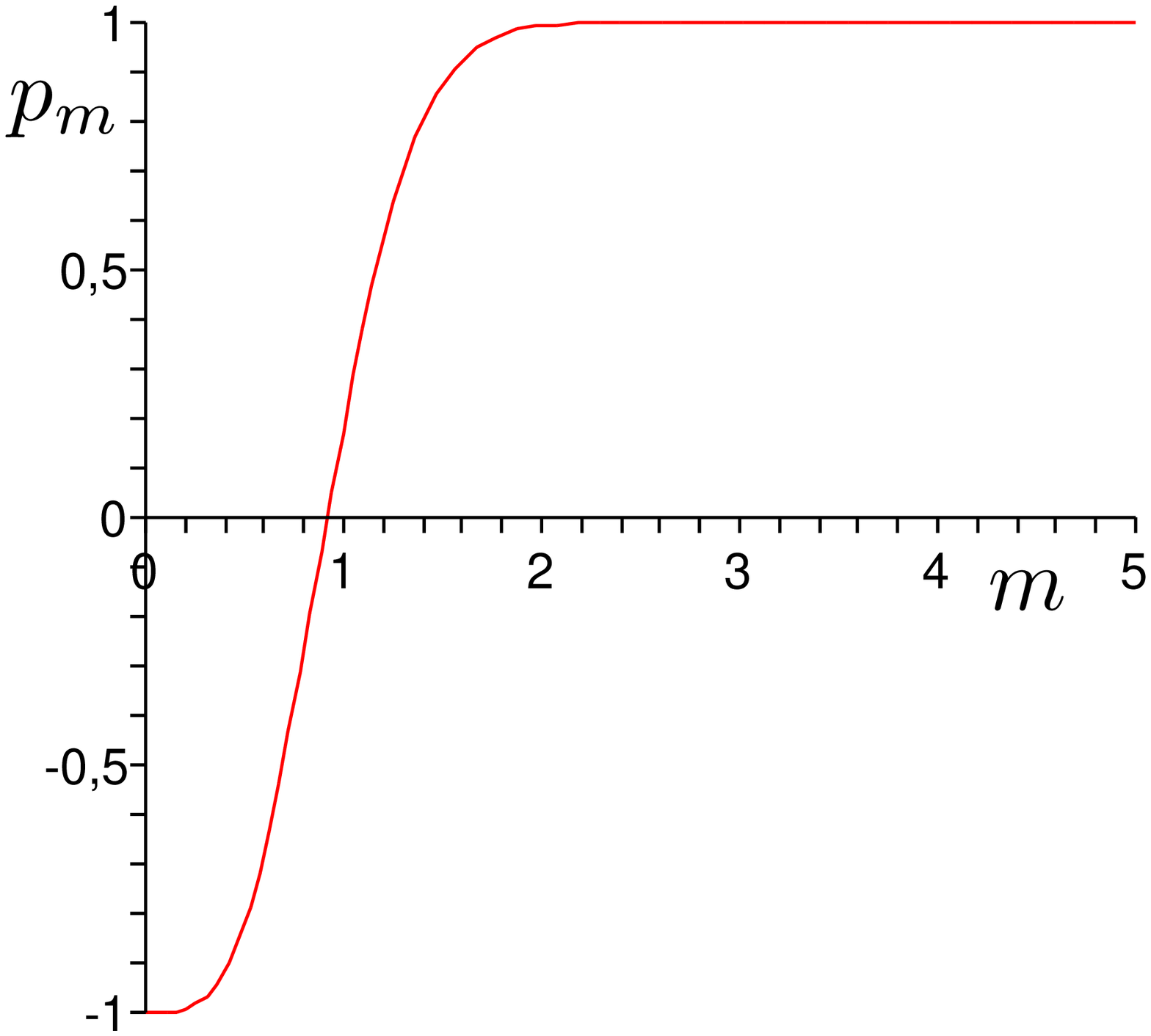}
\caption{ The dependence of $p_m$ on $m$ at $c=1/2$ (right), $c=1$
(center) and $c=2$ (left).} \label{eta_m}
\end{figure}

If and only if $c>1$, then there exists the interval of
$0<m^2<m^2_{max}$, on which $\eta_m<0$. Some part of this interval
is not physical, because $g(m^2,c)<0$ on this part. The
straightforward calculations (compare with~\cite{AJV0701184}) show
that at the point
\begin{equation}
m_{max}^2={}-\frac12-\frac12 W_{-1}\left(-\frac{e^{-1}}{c}\right),
\end{equation}
we obtain $\eta_m(m_{max})=0$. So, for $c>1$ and
$\xi^2<\xi^2_{max}$ we have two positive roots of (\ref{5c}):
$m_1$ and $m_2>m_1$, with $\eta_{m_1}<0$ and $\eta_{m_2}>0$. In
the next section we use this fact to construct a quintom local
model with one tachyon real scalar field, which corresponds to
$\eta_{m_2}$, and one phantom real scalar field, which corresponds
to $\eta_{m_1}$. For different values of $c$ the function
$p_m\equiv m^2\eta_m$ is presented in Fig.~\ref{eta_m}.

\section{Construction of solutions in the Friedmann--Robert\-son--Walker metric}

\subsection{Equations of motion and Friedmann equations}
In the spatially flat Friedmann--Robertson--Walker Universe we get
the following equation of motion for the space homogeneous scalar
field $\phi$
\begin{equation}
\label{equphiNFrid} (\xi^2\Dd +1) e^{-2 \Dd}\phi= c\:\phi.
\end{equation}

The Friedmann equations have the following form
\begin{equation}
\left\{
\begin{array}{l}
\displaystyle 3H^2= \frac{1}{m_p^2}~{\cal E},
\\[7.2mm]
\displaystyle \dot H={}-\frac{1}{2m_p^2}~({\cal E}+{\cal P}),
\end{array}
\right. \label{eomprho}
\end{equation}
where dot denotes the time derivative ($\dot H\equiv\partial_0
H$).

The second equation of system (\ref{eomprho}) is the nonlinear
integral equation in $H(t)$:
\begin{equation}
\dot H={}-\frac{1}{m_p^2}\left(
\frac{\xi^2}{2}(\partial_0\phi)^2-c\int\limits_{0}^{1}\!\left(\partial_0
e^{(1+\rho){\cal D}} \phi\right) \left(\partial_0 e^{(1-\rho)
{\cal D}}\phi\right)d\rho\right).
\end{equation}

Let us make an assumption, that $\phi(t)$ and $H(t)$ satisfy the
following equation
\begin{equation}
\label{equ1} \Dd \phi={}-\alpha^2\phi,
\end{equation}
where $\alpha$ is a root of eq.~(\ref{5c}).

In this case eq.~(\ref{equphiNFrid}) is solved. Using
formulas~(\ref{E1}) and (\ref{P1}), we rewrite
system~(\ref{eomprho}) in the following form:
\begin{equation}
\left\{
\begin{array}{l}
\displaystyle 3H^2=
\frac{\eta^{\vphantom{27}}_\alpha}{2m_p^2}\left(\dot\phi^2-
    \alpha^2\phi^2\right)+\frac{\Lambda'}{m_p^2},
\\[7.2mm]
\displaystyle \dot
H={}-\frac{\eta^{\vphantom{27}}_\alpha}{2m_p^2}\dot\phi^2.
\end{array}
\right. \label{eomprholocal1}
\end{equation}
It is easy to check that (\ref{equ1}) is a consequence of system
(\ref{eomprholocal1}). Instead of (\ref{eomprholocal1}) we can
consider the following third order system:
\begin{equation}
\left\{
\begin{array}{l}
\displaystyle \ddot\phi+3H\dot\phi=\alpha^2\phi,
\\[7.2mm]
\displaystyle \dot
H={}-\frac{\eta^{\vphantom{27}}_\alpha}{2m_p^2}\dot\phi^2.
\end{array}
\right. \label{eomprholocal1b}
\end{equation}
This system has the following integral of motion:
\begin{equation}
  I_1=3H^2-\frac{\eta^{\vphantom{27}}_\alpha}{2m_p^2}\left(\dot\phi^2-
    \alpha^2\phi^2\right)=\frac{1}{m_p^2}\Lambda',
\end{equation}
therefore, choosing the initial date for (\ref{eomprholocal1b})
one fixes the value of $\Lambda'$.

So, our assumption allows to transform a system with a nonlocal
scalar field into a system with a local one. In the same way we
obtain systems with two or more local fields. Let
\begin{equation}
\label{phiexpsum2}
    \phi(t)=\sum\limits_{n=1}^{N} \phi_n(t),
\end{equation}
where all $\phi_n(t)$ are solutions of (\ref{equ1}) with the same
function $H(t)$ and different values of $\alpha$:
$\alpha=\alpha_n$. If all $\alpha_n$ ($n=1..N$) are different
roots of (\ref{5c}), then system~(\ref{eomprho}) transforms into
the following system with $N$ scalar fields:
\begin{equation}
\left\{
\begin{array}{l}
\displaystyle 3H^2=
\frac{1}{2m_p^2}\left(\sum\limits_{n=1}^N
\eta^{\vphantom{27}}_{\alpha_n}\left(\dot\phi_n^2-
    \alpha_n^2\phi_n^2\right)+2\Lambda'\right),
\\[7.2mm]
\displaystyle \dot H={}-\frac{1}{2m_p^2}\left(\sum\limits_{n=1}^N
\eta^{\vphantom{27}}_{\alpha_n}\dot\phi_n^2\right).
\end{array}
\right. \label{eomprholocalN}
\end{equation}

In the case of two real roots $\alpha_1>0$ and
$\alpha_2>\alpha_1$:
\begin{equation}
\left\{
\begin{array}{l}
\displaystyle 3H^2=
\frac{1}{2m_p^2}\left(\eta^{\vphantom{27}}_{\alpha_1}\left(\dot\phi_1^2-
    \alpha_1^2\phi_1^2\right)+\eta^{\vphantom{27}}_{\alpha_2}\left(\dot\phi_2^2-
    \alpha_2^2\phi_2^2\right)+2\Lambda'\right),
\\[7.2mm]
\displaystyle \dot H={}-\frac{1}{2m_p^2}\left(
\eta^{\vphantom{27}}_{\alpha_1}\dot\phi_1^2+
\eta^{\vphantom{27}}_{\alpha_2}\dot\phi_2^2\right),
\end{array}
\right. \label{eomprholocal2}
\end{equation}
we have obtained that $\eta_{\alpha_1}<0$ and $\eta_{\alpha_2}>0$.
Therefore the corresponding two-field model is a quintom one, in
other words, includes  one phantom scalar field
($\eta_{\alpha_1}<0$ ) and one  scalar field  with the canonical
kinetic term ($\eta_{\alpha_2}>0$ ) and with the tachyon mass term
($\alpha_2^2\eta_{\alpha_2}>0$). The SFT inspired nonlinear local
quintom models and their exact solutions have been studied, for
example, in~\cite{AKVtwofields,Vtwofields}. To obtain exact
solutions with physically important properties usually one should
add some additional terms in the potential, which tend to zero in
the limit of the flat
space-time~\cite{AJV0701184,AKVtwofields,Vtwofields,AKV}. It is
interesting that system (\ref{eomprholocal1}) allows to find a
physically important exact solution without adding any term in the
potential.

\subsection{Exact Solution in the case $N=1$}

Let us consider system (\ref{eomprholocal1}) with real $\alpha$.
Two exact nontrivial real solutions of this system have been
presented in~\cite{AVzeta}. In our notations these solutions are
the following:

\begin{itemize}

\item At $\alpha\neq 0$ and $\eta^{\vphantom{27}}_\alpha<0$
\begin{equation}
  \phi(t)=A(t-t_0),\qquad \Lambda^\prime={}-A^2,
  \qquad H(t)=\frac{\alpha^2}{3}(t-t_0),
\end{equation}
where
\begin{equation}
A=\pm\sqrt{{}-\frac{2m_p^2\alpha^2}{3\eta^{\vphantom{27}}_\alpha}},
\end{equation}
$t_0$ is an arbitrary constant.

\item At $\alpha=0$, $\Lambda^\prime=0$ and
$\eta^{\vphantom{27}}_\alpha=\xi^2-2>0$
\begin{equation}
    \phi(t)=\pm\sqrt{\frac{2m_p^2}{3\eta^{\vphantom{27}}_\alpha}}\ln(t-t_0)+C_1,
    \qquad H(t)=\frac{1}{3(t-t_0)},
\end{equation}
where $t_0$ and $C_1$ are arbitrary constants. Note that the root
$\alpha=0$ exists if and only if $c=1$.

\end{itemize}

In this paper we present a new solution, which looks more
realistic for the SFT inspired cosmological model. At present time
one of the possible scenarios of the Universe evolution considers
the Universe to be a D3-brane (3 spatial and one time variable)
embedded in higher-dimensional space-time. This D-brane is
unstable and does evolve to the stable state. This process is
described by the dynamics of the open string, which ends are
attached to the brane (see reviews \cite{review-sft} and
references therein).
 A phantom scalar field is an open string theory tachyon.
According to the Sen's conjecture~\cite{Sen-g} this tachyon
describes brane decay, at which a slow transition in a stable
vacuum takes place. This vacuum is characterized by the absence of
open string states, \textit{i.e.} corresponds to states of the
closed string. This picture allows us to specify the asymptotic
conditions for the scalar field. We assume that the phantom field
$\phi(t)$ smoothly rolls from the unstable perturbative vacuum
($\phi=0$) to a nonperturbative one, for example $\phi=A_0$, where
$A_0$ is a nonzero constant, and stops there. It is easy to see
that exact solutions, presented in~\cite{AVzeta} do not satisfy
these conditions.

 At $c=1$ our model~(\ref{action-1}) is a nonlocal model for the dilaton
coupling to the gravitation field. Its distinguished feature is
the invariance under the shift of the dilaton field to a constant.
In this case one of solutions of eq.~(\ref{5c}) is $\alpha=0$.
Summing the first and the second equations of
(\ref{eomprholocal1}), we obtain:
\begin{equation}
\dot H=\frac{\Lambda^\prime}{m_p^2}-3H^2.
\end{equation}
If $\Lambda^\prime>0$, then we obtain a real solution:
\begin{equation}
\label{Hnew}
H_1(t)=\sqrt{\frac{\Lambda^\prime}{3m_p^2}}\tanh\left(\sqrt{\frac{3\Lambda^\prime}
{m_p^2}}(t-t_0)\right),
\end{equation}
where $t_0$ is an arbitrary real constant.

\begin{figure}[h]
\centering
\includegraphics[width=72mm]{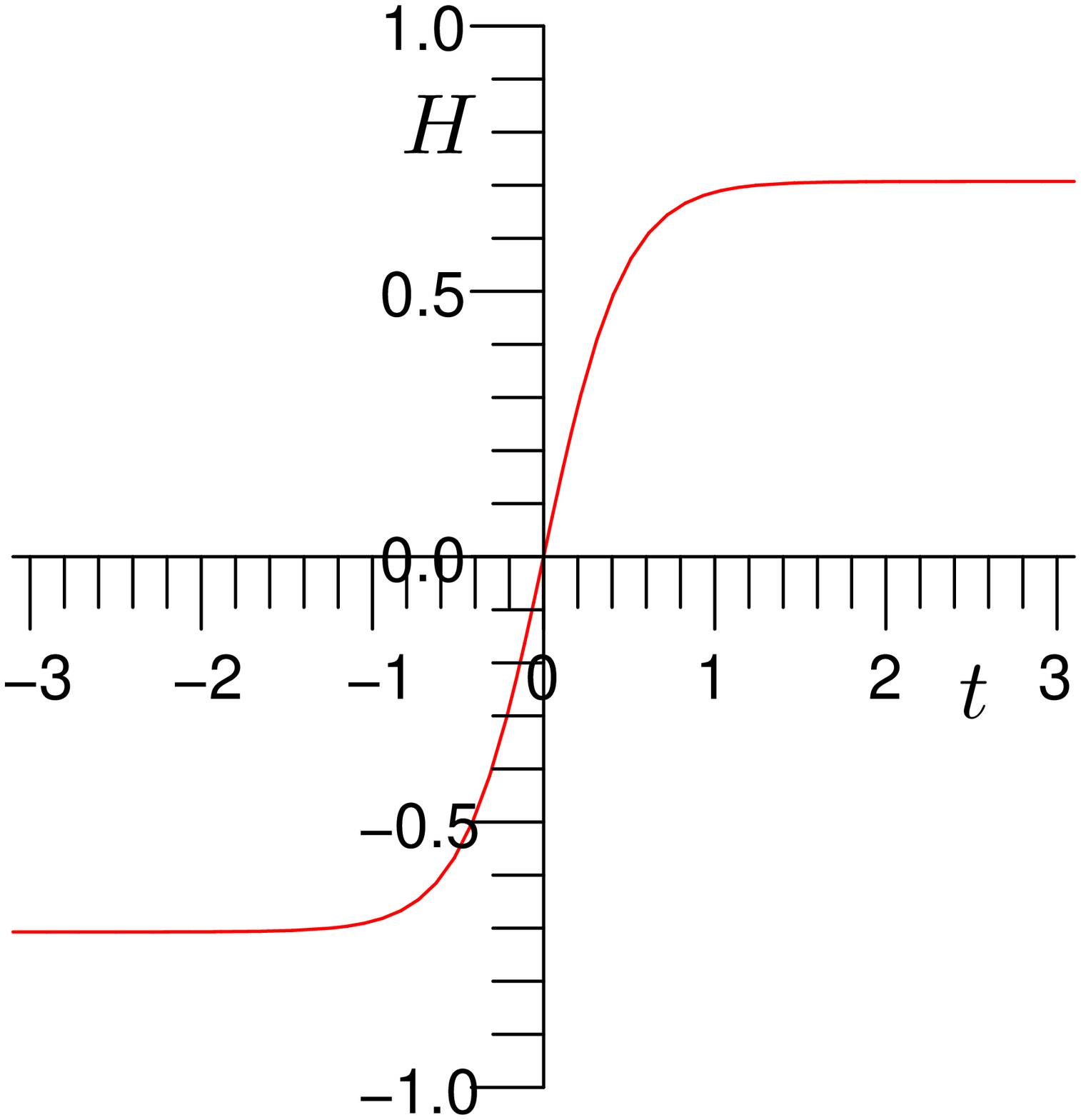} \ \ \ \ \
\includegraphics[width=72mm]{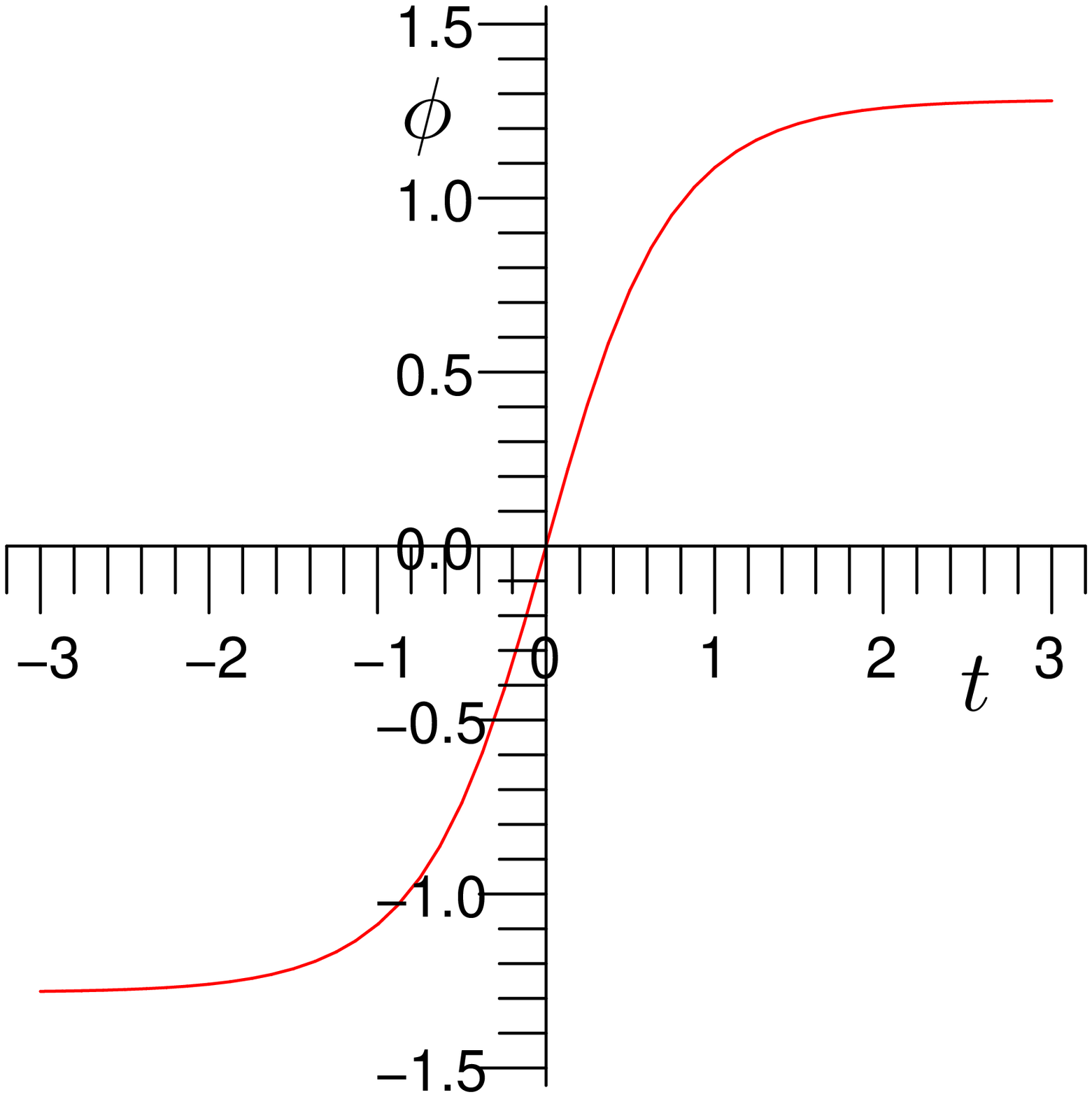}
\caption{ The functions $H_1(t)$  (right) and $\phi_1(t)$ (left)
at $\Lambda'=\frac{3}{2}$, $m_p^2=1$, $\xi^2=1$, $t_0=0$ and
$C_2=0$.} \label{H1phi1}
\end{figure}

It is easy to see that $\dot H_1(t)>0$ for any $t$, hence, from
the second equation of (\ref{eomprholocal1}) we obtain that
$\phi(t)$ can be real scalar field only if it is a phantom one
($\eta_\alpha<0$, that is equivalent to $\xi^2<2$). The explicit
form of $\phi(t)$ is as follows:
\begin{equation}
\phi_1(t)=\pm\sqrt{\frac{2m_p^2}{3(2-\xi^2)}}\arctan
\left(\sinh\left(\sqrt{\frac{3\Lambda^\prime}{m_p^2}}(t-t_0)\right)\right)+C_2,
\end{equation}
where $C_2$ is an arbitrary constant. Functions $H_1(t)$ and
$\phi_1(t)$ are presented in Fig.~\ref{H1phi1}.

The Hubble parameter $H_1(t)$ is a monotonically increasing
function, so, using that
\begin{equation}
\label{w} w={}-1-\frac23\frac{\dot H_1}{H_1^2},
\end{equation}
we obtain $w<-1$. So, solution (\ref{Hnew}) corresponds to phantom
dark energy. Note that we have found two-parameter set of exact
solutions at any $\Lambda^\prime>0$. In other words, at any
$\Lambda^\prime>0$ we have found the general solution
of~(\ref{eomprholocal1}), which correspond to $\alpha=0$. At
$\Lambda^\prime=0$ the solution (\ref{Hnew}) transforms to a
constant. In the case~$\Lambda^\prime=0$ the general solution has
been found in~\cite{AVzeta}.

In the case $\Lambda^\prime<0$ we obtain the following general
solution:
\begin{equation}
H_2(t)={}-\sqrt{\frac{-{}\Lambda^\prime}{3m_p^2}}\tan\left(\sqrt{{}
-\frac{3\Lambda^\prime}{m_p^2}}(t-t_0)\right),
\end{equation}

\begin{equation}
\phi_2(t)={}\pm \sqrt{\frac{8m_p^2}{3(\xi^2-2)}}
\mathrm{arctanh}\left(\frac{\cos\left(\sqrt{\frac{{}-3\Lambda^\prime}
{m_p^2}}(t-t_0)\right)-1}{\sin\left(\sqrt{\frac{{}-3\Lambda^\prime}
{m_p^2}}(t-t_0)\right)}\right)+C_2.
\end{equation}

This solution is real at $\xi^2>2$. It is interesting that the
type of solutions essentially depends on sign of $\Lambda^\prime$.
The solution with the SFT inspired boundary conditions corresponds
to $\Lambda^\prime>0$.

\section{Conclusions}

We have studied the SFT inspired linear nonlocal model. This model
has an infinite number of higher derivative terms and are
characterized by two positive parameters: $\xi^2$ and $c$. For
particular cases of the parameters $\xi^2$ and $c$ the
corresponding actions describe linear approximations to either the
bosonic or nonBPS fermionic cubic SFT as well as  to the
nonpolynomial SFT.

Roots of the characteristic equation do not depend on the form of
the metric and this property allows us to study properties of
energy density and pressure. We have found that in an arbitrary
metric the energy-momentum tensor for an arbitrary N-mode solution
is a sum of the energy-momentum tensors for the corresponding
one-mode solutions. In the Friedmann--Robertson--Walker spatially
flat metric the pressure for a one-mode solution corresponding to
a real root can be positive or negative, depending on parameters
of our nonlocal model. Namely, for $c\leq 1$ the one mode pressure
is positive and for $c>1$ it could be negative or positive.

The investigation performed in this paper shows that the general
field equations in  linear nonlocal  models admit an equivalent
description in terms of local theory and as a consequences we have
representations (37) and (38) for the energy and pressure. This
calculation also supports the use of the Ostrogradski
representation for our system in the case of arbitrary metric.

In distinguish to our previous paper~\cite{AJV0701184} we do not
use any approximation scheme and do not add any terms in the
potential. We have shown that our linear model with one nonlocal
scalar field generates an infinite number of local models. These
models can be studied numerically and we plan to present this
analysis in future papers. Some of these models have been solved
explicitly and, hence, special exact solutions for nonlocal model
in the Friedmann--Robertson--Walker metric have been obtained. In
particular we have constructed an exact kink-like solution, which
correspond to monotonically increasing Universe with phantom dark
energy. Note that the obtained behaviour of the Hubble parameter
is close to behavior of the Hubble parameter in the nonlinear
nonlocal model~\cite{IA1}, which recently has been obtained
numerically~\cite{Jukovskaya0707}.

\section*{Acknowledgement}
L.J. would like to thank D. Mulryne for fruitful discussions.
 S.V. is grateful to the organizers of
\href{http://tristan.fam.cie.uva.es/~qts5/}{the V-th International
Symposium "Quantum Theory and Symmetries"}
(\href{http://tristan.fam.cie.uva.es/~qts5/}{QTS'5}) for
hospitality and financial support (the grant awarded by the
Fundacion Universidades de Castilla y Leon), his participation in
the QTS'5 conference has been supported in part by the European
Physical Society and RFBR travel grant 07-01-08245.

This research is supported in part by RFBR grant 05-01-00758.
The work of I.A. and L.J. is supported in part by INTAS grant 03-51-6346
and Russian President's grant NSh--672.2006.1.  S.V.
is supported in part by Russian President's grant NSh--8122.2006.2.
L.J. acknowledges the
support of the Centre for Theoretical Cosmology, in Cambridge.


\end{document}